\documentclass[11pt, a4paper]{article}
\usepackage{moriond,epsfig}

\def\be{\begin{equation}}
\def\ee{\end{equation}}
\def\bea{\begin{eqnarray}}
\def\eea{\end{eqnarray}}

\def\pt{p_t}
\def\as{\alpha_s}
\def\ie{i.\ e.\ }

\begin{document}
\vspace*{4cm}
\title{Accurate predictions for heavy quark jets}

\author{Giulia Zanderighi(1)}
\address{(1) CERN, Geneve 23, CH-1211, Switzerland}

\maketitle\abstracts{
Heavy-flavour jets enter many of today's collider studies, yet NLO
predictions for these quantities are subject to large
uncertainties, larger than the corresponding experimental errors. We
propose a new, infrared safe definition of heavy-quark jets which
allows one to reduce theoretical uncertainties by a factor of three. 
}

\section{Introduction}
  \begin{figure}[b]
  \centering
\psfig{figure=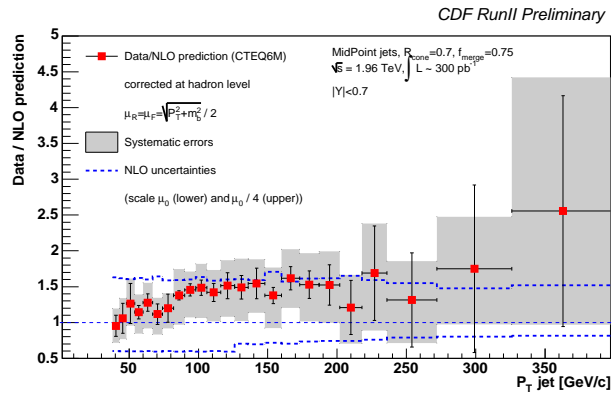,width=7.9cm}
\caption{Ratio of the measured inclusive $b$-jet spectrum to the NLO prediction. The measurement is performed for jets with transverse momentum 38
  GeV $< P_{T,\rm jet} < $ 400 GeV and rapidity $|y_{\rm jet}| <$
    0.7.
}
  \label{fig:cdf-bjets}
\end{figure}
\begin{figure}[t]
  \centering
\psfig{figure=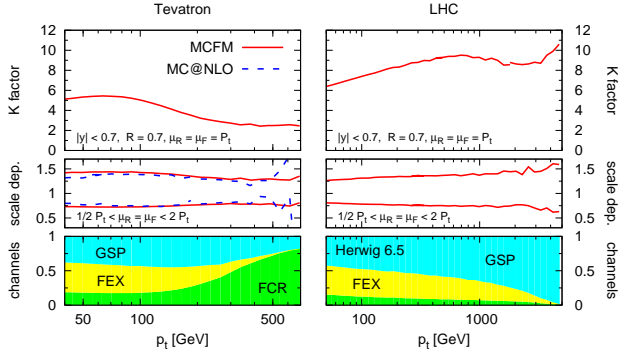,width=4.6cm,angle=270}
\caption{Top: $K$-factor for inclusive $b$-jet spectrum as computed with 
    MCFM, clustering particles into jets using the $k_t$
    jet-algorithm with $R$=0.7, and selecting jets in the central
    rapidity region ($|y| <0.7$). Middle: scale dependence obtained by
    simultaneously varying the renormalisation and factorisation
    scales by a factor two around $\pt$, the transverse momentum of
    the hardest jet in the event. Bottom: breakdown of the Herwig inclusive
    $b$-jet spectrum into the three major hard underlying channels
    contributions (for simplicity the small $bb \to bb$ is not
    shown).
}
  \label{fig:kfact+channels}
\end{figure}

 When looking at the current comparison between the
inclusive $b$-jet spectra measured by CDF and the corresponding
next-to-leading order (NLO)
predictions, Fig.~\ref{fig:cdf-bjets}~\cite{cdf-bjets},
 one notices two striking features.
Firstly, one sees a tension between data and theory: the ratio of data
over NLO is around 1.2-1.5 over the whole range of accessible
transverse momenta $p_t$ of the jets.  Secondly, one notices that the
uncertainties associated with the theoretical predictions are
embarrassingly large ($\sim 40-50$\%) for a NLO calculation and in
particular they are larger than the corresponding experimental
uncertainties. To understand why this happens it is useful to examine
Fig.~\ref{fig:kfact+channels}.

The top plots show that the large uncertainty is associated with very
large $K$-factors. The middle plots confirm that the uncertainty is
the same both with MCFM~\cite{mcfm} and
MC\@NLO~\cite{Frixione:2003ei}. Finally, the bottom plots illustrate
the origin of the poor convergence of the perturbative expansion: when
breaking down the Herwig~\cite{Herwig} $b$-jet spectrum into the hard
underlying channels it turns out that two NLO channels, flavour
excitation, where a $b$-quark is kicked out of the sea-quarks, and
gluon splitting, where a gluon in the final state splits into a $b\bar
b$-pair, are larger than the leading order heavy quark production
mechanism, flavour creation, when two incoming light partons produce a
heavy quark pair.

The reason why supposedly higher order contributions are actually
larger than the leading order channel can be clarified by counting
soft and collinear logarithms associated with the splitting of gluons
into $b\bar b$-pairs. It turns out that flavour excitation contributes
with $(\as \ln \pt/m_b)^n$ and gluon splitting contributes with $(\as
\ln \pt/m_b)^{2n-1}$ relative to the leading order, ${\cal O}(\as^2)$
process. Since $m_b\ll \pt$ these contributions are enhanced.
Moreover, the dominant contribution to the b-jet spectrum comes from
jets originated from gluon splitting, which do not correspond to one's
intuitive physical idea of a b-jet, one where a hard $b$ is produced
directly in the hard scattering.~\footnote{We recall that according
to the current experimental definition of a $b$-jets, a $b$-jet is any
jet containing {\em at least} one $b$.}  In the following we suggest
to adopt a different jet-clustering algorithm to reconstruct
$b$-jets. One that by making explicit use of the flavour information
eliminates all higher-order logarithmic enhancements associated to
gluon splittings in the $b$-jet spectra.  This means that, after
resumming initial state collinear logarithms into $b$-pdfs, $b$-jets
can be computed using massless QCD calculations~\cite{NLOJET} as long
as one neglects power corrections $m_b^2/p_t^2$ (potentially
log-enhanced).

\section{The heavy-quark jet algorithm}
\label{sec:algo}
We summarize here the inclusive heavy-flavour jet
algorithm for hadron-hadron collisions~\cite{jetflav}.
For any pair of final-state particles $i$, $j$ define a class of
  distances $d_{ij}^{(F,\alpha)}$
  parametrized by $0<\alpha\le 2$ and a jet radius $R$
  \begin{equation}
    \label{eq:dij-flavour-alpha}
    d_{ij}^{(F,\alpha)} = \frac{R_{ij}^2}{R^2}
    \times\left\{ 
      \begin{array}[c]{ll}
        \max(k_{ti}, k_{tj})^\alpha \min(k_{ti}, k_{tj})^{2-\alpha}\,,
        & \mbox{softer of $i,j$ flavoured,}\\
        \min(k_{ti}^2, k_{tj}^2)\,, &  
        \mbox{softer of $i,j$ flavourless,}
      \end{array}
    \right.
  \end{equation}
where $R_{ij}^2 = \Delta y_{ij}^2 + \Delta \phi_{ij}^2$, $\Delta
y_{ij} = y_i - y_j$, $\Delta\phi_{ij} = \phi_i -
  \phi_j$ and $k_{ti}$, $ y_i$ and $\phi_i$ are respectively the
  transverse momentum, rapidity and azimuth of particle $i$, with
  respect to the beam.
  For each particle $i$ define a distance with respect to the beam $B$ at
  positive rapidity,
  \begin{equation}
    \label{eq:diB-flavour-alpha}
    d_{iB}^{(F,\alpha)} = \left\{
      \begin{array}[c]{ll}
        \max(k_{ti}, k_{tB}( y_i))^\alpha 
        \min(k_{ti}, k_{tB}( y_i))^{2-\alpha}
        \,, & \quad\mbox{$i$ is flavoured,}\\
        \min(k_{ti}^2, k_{tB}^2( y_i))\,, & \quad\mbox{$i$ is flavourless,}
      \end{array}
    \right.
  \end{equation}
with 
\begin{equation}
  \label{eq:ktB}
  k_{tB} ( y) =
  \sum_i k_{ti} \left( \Theta( y_i -  y) +
    \Theta( y -  y_i) e^{ y_i -  y}\right)\,.
\end{equation} 
Similarly define a distance to the beam $\bar B$ at negative rapidity
by replacing $k_{tB}$ in eq.~(\ref{eq:diB-flavour-alpha}) with $k_{t
  \bar B}$
\begin{equation}
  \label{eq:ktBbar}
  k_{t\bar B} ( y) = 
  \sum_i k_{ti} \left( \Theta( y -  y_i) +
    \Theta( y_i -  y) e^{ y -  y_i}\right)\,.
\end{equation}
Identify the smallest of the distance measures. If it is a
  $d_{ij}^{(F,\alpha)}$, recombine $i$ and $j$ into a new particle,
  summing their flavours and 4-momenta; if it is a
  $d_{iB}^{(F,\alpha)}$ (or $d_{i\bar B}^{(F,\alpha)}$) declare $i$ to be
  a jet and remove it from the list of particles.
Repeat the procedure until no particles are left.
We define the $b$-flavour or generally the heavy-flavour
of a (pseudo)-particle or a jet as its net heavy flavour
content, \ie the total number of heavy quarks minus heavy anti-quarks.

The IR-safety of this algorithm was proved in~\cite{jetflav}. Apart
from allowing one to take the limit $m_Q^2\to 0$ for the heavy quark mass (as long as collinear
singularities associated with incoming heavy quarks are factorized
into a heavy quark PDF),
it ensures that one
obtains the same results whether one considers heavy-quark flavour at
parton level, or heavy-meson flavour at hadron level, modulo
corrections suppressed by powers of $\Lambda_{QCD}/\pt$.   

\section{Results} 
\begin{figure}[t]
  \centering
  \psfig{figure=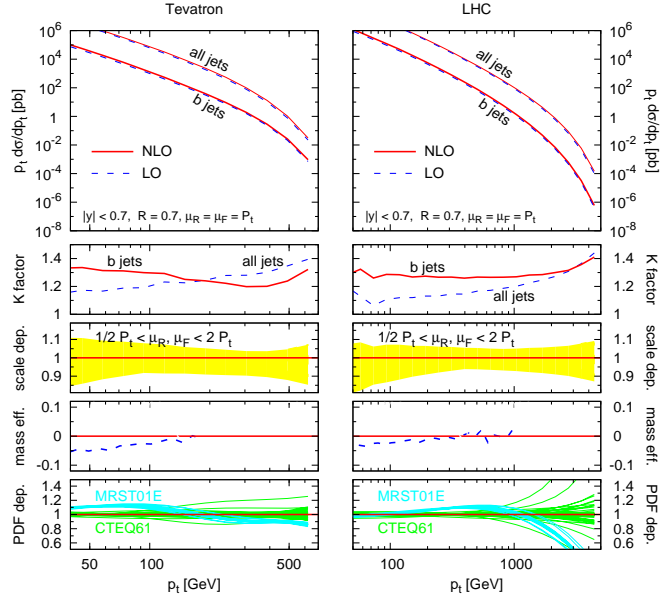,width=7.6cm,angle=270}
\caption{Inclusive jet spectrum at the Tevatron (right) and at the
    LHC (left). The top two panels show results for both $b$-jets and
    all-jets, while the lower three panels apply only to $b$-jets.
    See text for further details.}
  \label{fig:two-spect}
\end{figure}

Our results are summarized in
fig.~\ref{fig:two-spect}~\cite{Banfi:2007gu} where we show the
inclusive $b$-jet $\pt$-spectrum as obtained with the flavour
algorithm specified above with $\alpha=1$, and $R=0.7$, the latter
having been shown to limit corrections associated with the
non-perturbative underlying event~\cite{cdf-jets}. The left (right)
column of the figure shows results for the Tevatron Run II (LHC). We
have selected only those jets with rapidity $|y| <0.7$. We also show
the full inclusive jet spectrum (all jets) as obtained with a standard
inclusive $k_t$-algorithm~\cite{kthh} with $R=0.7$.

We notice the considerable reduction of $K$-factors, which are around
1.3 and the moderate uncertainties associated with scale variation,
    signaling
that the perturbative expansion is now well under control.  Our
predictions constitute therefore the first accurate predictions for
inclusive heavy quark jets.  

We remark that very similar results are obtained when considering
charmed jet spectra.  An interesting issue there is that predictions
are very sensitive to possible intrinsic charm components of the
proton~\cite{Pumplin:2007wg}. This means that this type of observable
has a potential to set constraints on such intrinsic components.

A last remark concerns the feasibility of the experimental measurement
of heavy flavour jets defined with our flavour algorithm. Our
jet-clustering algorithm requires that one identify heavy-flavoured
particles and that one uses a different distance measure when
clustering heavy or light objects according to
eq.~(\ref{eq:dij-flavour-alpha}).  It is particularly important to
identify cases when both heavy flavoured particles are in the same
jet, so as to label this jet a gluon jet and eliminate it from the
$b$-jet spectrum.  Experimentally techniques for double $b$-tagging in
the same jet already exist ~\cite{Acosta:2004nj} and steady progress
is to be expected in the near future~\cite{btag}.  However one has
always a limited efficiency for single $b$ tagging, and even more for
double $b$-tagging in the same jet.  On the other hand preliminary
studies indicate that one does not necessarily need high efficiencies,
but what is more crucial is that one understand those efficiencies
well~\cite{Banfi:2007gu}. We look forward to further investigation in
this direction.

\section*{Acknowledgments}
This work is done in collaboration with Andrea Banfi and Gavin Salam.


\end{document}